\documentclass[11pt]{article}
\usepackage{amsmath, amsthm, amssymb,color,fullpage,url,booktabs}  
\usepackage{graphicx}
\ifdefined\blackandwhite
\usepackage[pdftex]{hyperref} 
\else
\usepackage[colorlinks,pdftex]{hyperref} 
\hypersetup{citecolor = blue, linkcolor = red!70!black}
\fi

\usepackage{cleveref}
\usepackage{tikz}

\newcommand{\nc}{\newcommand}
\nc{\rnc}{\renewcommand}

\newcommand{\bra}[1]{\left\langle #1\right|}
\newcommand{\ket}[1]{\left|#1\right\rangle}
\newcommand{\proj}[1]{\left|#1\right\rangle\left\langle #1\right|}
\newcommand{\braket}[2]{\left\langle #1\middle|#2\right\rangle}

\DeclareMathOperator{\poly}{poly}

\DeclareMathOperator{\tr}{tr}

\DeclareMathOperator{\BPP}{\mathsf{BPP}}
\DeclareMathOperator{\BQP}{\mathsf{BQP}}

\DeclareMathOperator{\NP}{\mathsf{NP}}

\DeclareMathOperator{\PH}{\mathsf{PH}}
\DeclareMathOperator{\SharpP}{\mathsf{\#P}}

\DeclareMathOperator{\Ptime}{\mathsf{P}}

\DeclareMathOperator{\PostBPP}{\mathsf{PostBPP}}
\DeclareMathOperator{\PostBQP}{\mathsf{PostBQP}}

\def\be#1\ee{\begin{equation}#1\end{equation}}
\def\ba#1\ea{\begin{align}#1\end{align}}
\def\bas#1\eas{\begin{align*}#1\end{align*}}
\def\bpm#1\epm{\begin{pmatrix}#1\end{pmatrix}}
\nc{\non}{\nonumber}
\nc{\nn}{\nonumber}
\nc{\eq}[1]{(\ref{eq:#1})}
\nc{\eqs}[2]{(\ref{eq:#1}) and (\ref{eq:#2})}
\rnc{\L}{\left} 
\nc{\R}{\right}
\nc{\ra}{\rightarrow}
\nc{\ot}{\otimes}
\nc{\grad}{{\vec{\nabla}}}

\def\bea#1\eea{\begin{eqnarray}#1\end{eqnarray}}
\def\beas#1\eeas{\begin{eqnarray*}#1\end{eqnarray*}}

\newtheorem{thm}{Theorem}
\newtheorem*{thm*}{Theorem}

\newtheorem{proto}{Protocol}

\theoremstyle{definition}

\newtheorem{dfn}[thm]{Definition}
\theoremstyle{plain}

\makeatletter
\newtheorem*{rep@theorem}{\rep@title}
\newcommand{\newreptheorem}[2]{%
\newenvironment{rep#1}[1]{%
 \def\rep@title{#2 \ref{##1} (restatement)}%
 \begin{rep@theorem}}%
 {\end{rep@theorem}}}
\makeatother

\newreptheorem{thm}{Theorem}
\newreptheorem{lem}{Lemma}

\nc\eps{\epsilon}

\nc\cA{\mathcal{A}}
\nc\cB{\mathcal{B}}
\nc\cC{\mathcal{C}}
\nc\cD{\mathcal{D}}
\nc\cE{\mathcal{E}}
\nc\cF{\mathcal{F}}
\nc\cG{\mathcal{G}}
\nc\cH{\mathcal{H}}
\nc\cI{\mathcal{I}}
\nc\cJ{\mathcal{J}}
\nc\cK{\mathcal{K}}
\nc\cL{\mathcal{L}}
\nc\cM{\mathcal{M}}
\nc\cN{\mathcal{N}}
\nc\cO{\mathcal{O}}
\nc\cP{\mathcal{P}}
\nc\cQ{\mathcal{Q}}
\nc\cR{\mathcal{R}}
\nc\cS{\mathcal{S}}
\nc\cT{\mathcal{T}}
\nc\cU{\mathcal{U}}
\nc\cV{\mathcal{V}}
\nc\cW{\mathcal{W}}
\nc\cX{\mathcal{X}}
\nc\cY{\mathcal{Y}}
\nc\cZ{\mathcal{Z}}

\nc\bbC{\mathbb{C}}

\nc\bbF{\mathbb{F}}
\nc\bbM{\mathbb{M}}
\nc\bbN{\mathbb{N}}
\nc\bbR{\mathbb{R}}
\nc\bbZ{\mathbb{Z}}

\nc\benum{\begin{enumerate}}
\nc\eenum{\end{enumerate}}
\nc\bit{\begin{itemize}}
\nc\eit{\end{itemize}}
\newcommand{\fig}[1]{Fig.~\ref{fig:#1}}
\newcommand{\secref}[1]{Section~\ref{sec:#1}}

\nc{\todo}[1]{\textcolor{red}{todo: #1}}
\nc{\Anote}[1]{\textcolor{red}{Aram note: #1}}

\def\begsub#1#2\endsub{\begin{subequations}\label{eq:#1}\begin{align}#2\end{align}\end{subequations}}
\nc\qand{\qquad\text{and}\qquad}
\nc\mnb[1]{\medskip\noindent{\bf #1}}

\nc{\pder}[2]{\frac{\partial {#1}}{\partial {#2}}}
\nc{\p}{\partial}
\usepackage{tikz}
\usetikzlibrary{backgrounds,fit,decorations.pathreplacing,calc}
\usepackage{tabularx}
\usepackage{cite}

\DeclareMathOperator{\eval}{Check}
\DeclareMathOperator{\PostQAOA}{\mathsf{PostQAOA}}

\renewcommand{\bra}[1]{\langle #1|}
\renewcommand{\ket}[1]{|#1\rangle}

\newcommand{\Ket}[1]{\left|#1\right\rangle}

\newcommand{\expb}[1]{\exp\bigl( #1 \bigr)}
\newcommand{\expp}[1]{\exp( #1)}
\renewcommand{\Ket}[1]{\ensuremath{\left|#1\right\rangle}}

\newcommand\QAdi{QADI}
\newcommand\QAdiS{\QAdi-SG}
\newcommand\QApx{QAOA}

\begin{document}

\title{Quantum Supremacy through the Quantum Approximate Optimization Algorithm}

\author{Edward Farhi\thanks{Google Inc.}
\thanks{Center for Theoretical
    Physics.  MIT} \and Aram W. Harrow\footnotemark[2]}
\maketitle
\begin{abstract}
The Quantum Approximate Optimization Algorithm (QAOA) is designed to
run on a gate model quantum computer and has shallow depth.  It takes
as input a combinatorial optimization problem and outputs a string
that satisfies a high fraction of the maximum number of clauses that
can be satisfied.  For certain problems the lowest depth version of
the QAOA has provable performance guarantees although there exist
classical algorithms that have better guarantees.  Here we argue that
beyond its possible computational value the QAOA can exhibit a form of
``Quantum Supremacy'' in that, based on reasonable complexity theoretic
assumptions, the output distribution of even the lowest depth version
cannot be efficiently simulated on any classical device.  We contrast
this with the case of sampling from the output of a quantum computer
running the Quantum Adiabatic Algorithm (QADI) with the restriction
that the Hamiltonian that governs the evolution is gapped and
stoquastic.  Here we show that there is an oracle that would allow
sampling from the QADI but even with this oracle, if one could
efficiently classically sample from the output of the QAOA, the
Polynomial Hierarchy would collapse.  This suggests that the QAOA is
an excellent candidate to run on near term quantum computers not only
because it may be of use for optimization but also because of its
potential as a route to
establishing Quantum Supremacy. 
\end{abstract}

\section{Introduction}\label{sec:intro}
Feynman's original motivation for exploring quantum computing was that quantum
mechanics is hard to simulate on a classical computer.  In particular,
writing down all the amplitudes of a generic $n$-qubit state requires
$2^n$ complex numbers, which is far too large for any possible
classical computer say when $n \gtrsim 100$.  But what if you only want to
calculate a single amplitude of a large quantum state?  Or to sample
on a classical computer (with a random number generator) from the
distribution that would result from performing a quantum measurement?
The tasks of computing amplitudes or sampling can be viewed as
forms of simulation which might be feasible on
classical computers.
These tasks can sometimes be accomplished efficiently, using
problem-specific techniques.  For example, with the known Hydrogen atom
eigenstates, we can calculate matrix elements of operators of interest
to arbitrary accuracy.  

But there are no known methods for computing
amplitudes or sampling which run in less than exponential time on a
classical computer and work for all quantum systems.
  As small-scale quantum computers come online, we
want to understand how difficult it is to simulate their behavior on
classical computers.  For a such a device to be computationally
useful---that is, able to outperform a classical computer for a given
task---it had better be hard to classically simulate.

Complexity theoretic assumptions can be used to show that the tasks of classically computing matrix elements of a
general quantum circuit~\cite{FGHP} or classically sampling from the output of a
general quantum circuit~\cite{TD04} cannot be done efficiently.  In this paper we will review these arguments.  We
do not reproduce all of the details needed to establish these results
but give what we hope is a faithful outline of the arguments.   Related hardness results are known for particular families of quantum circuits such as Boson Sampling, small depth quantum circuits and IQP or Instantaneous Quantum Polytime circuits~\cite{AA13,BJS10,BMS15,Brod15,Saeed16,TD04}.

The Quantum
Approximate Optimization Algorithm~\cite{QAOA,QAOA2} or \QApx{} is a family of quantum circuits designed to find ``good''
solutions to optimization problems.  Here we will show, using methods
that are similar to those used in the IQP case, that classically
sampling from even the shallowest depth version of the \QApx{} can be
argued to be difficult for complexity theoretic reasons.  In this
sense the \QApx{} can exhibit a form of  ``Quantum Supremacy''~\cite{supremacy}, that is be a quantum process whose output distribution cannot be efficiently reproduced on a classical device.
It is also possible that versions of the \QApx{} can find approximate solutions to combinatorial search problems faster than classical algorithms.  This may be established by running the algorithms on actual devices or by theoretical analysis.  This possible algorithm advantage, together with our non-simulatability results, suggests that the \QApx{} is a good choice of algorithm to run on a near-term quantum computer.

We also
discuss the prospects of sampling from the evolving state of a system
running the Quantum Adiabatic Algorithm~\cite{farhi00,Nishimori:98a} which we call \QAdi{} (so as
to not use an acronym which is a subset of \QApx{}).  Here practitioners use numerical simulations of \QAdi{} which rely on the assumption that 
the Hamiltonian governing the evolution is stoquastic (see \secref{adiabatic} for definitions).  We will show that this assumption rules out using the same arguments for Quantum Supremacy that worked for \QApx{}.

Our paper is organized as follows.  First we review the basic
ingredients of the \QApx{}.  Since we are going to use complexity theory
arguments to argue that it is difficult to simulate this algorithm, we
review some of the needed complexity theory.  We explain what the
Polynomial Hierarchy (PH) is and what it means for it to collapse.  In \secref{mat-elt}, we
 show why the ability to compute on a classical computer the
matrix elements of the \QApx{} would collapse the PH.  But to discuss why
sampling the output distribution of the \QApx{} is also hard for
complexity theory reasons we need to discuss Post-Selected Quantum
Computing which we do in \secref{PostBQP}.  Using Post-Selected Quantum
Computing as a tool, in \secref{arb-collapse} we show that the efficient
sampling of the output of an arbitrary quantum circuit implies the
collapse of the PH.  This leads up to our new result (in Sections~\ref{sec:post-equal} and \ref{sec:QAOA-collapse}) that efficient sampling from the output of
the \QApx{} also collapses the PH.  In \secref{adiabatic} we turn to a discussion of
sampling from the output of the Quantum Adiabatic Algorithm being run
in optimization mode.  We review the assumptions needed to enable
efficient sampling from this type of quantum circuit.  Here we argue
that this may in fact be easier than sampling from the output of the
\QApx{}.    We conclude with a discussion of what these results mean for
near-term quantum computers.

\section{Background}
\subsection{Constraint satisfaction problems}
\label{sec:QA-CSP}
A constraint satisfaction problem (CSP) is specified by $n$ bits and a
collection of $m$ constraints (or clauses), each of which involves a small subset
of the bits.  The computational task is to find a string which
maximizes or approximately maximizes the number of satisfied
constraints. For each constraint $a \in [m]$ and each string $z\in
\{0,1\}^n$, define
\be
C_a(z) = \begin{cases} 1 & \text{if }z\text{ satisfies the constraint
  }a\\
0 & \text{if $z$ does not .}
\end{cases}\ee
Equivalently, the goal is to maximize
\be C(z) = \sum_{a=1}^m C_a(z),\label{eq:CSP-def}\ee
which counts the number of satisfied constraints.  One example of a
CSP is MAX-CUT, where the constraints are indexed by edges $\langle
i,j\rangle$ in a graph and are satisfied when $z_i$ and $z_j$
disagree; i.e.
\be C_{\langle i,j\rangle}(z) = \left( z_i-z_j \right)^2 .
\label{eq:disagree}\ee
Thus maximizing the sum of these clauses is equivalent to finding the
minimum energy of a Ising antiferromagnet where the spins sit on the
vertices of a graph.

In this paper we consider two quantum algorithms for CSPs: the Quantum
Approximate Optimization Algorithm and the Quantum Adiabatic
Algorithm.   We will come back to the second one later in the paper.
Both
operate in the $2^n$-dimensional Hilbert space with basis vectors $\ket{z}$
and accordingly we define the operator $C$ by
\be C \ket z := C(z) \ket z.\label{eq:C-def}\ee
We will also define the operator
\be B  := \sum_{i=1}^n \sigma_x^{(i)}, \label{eq:B-def}\ee
where $\sigma_\alpha^{(i)}$ denotes the Pauli operator $\sigma_\alpha$
acting on the $i^{\text{th}}$ qubit.
Both algorithms also start with
the same initial state
\be \ket{s} := \frac{1}{\sqrt{2^n}}\sum_{z\in \{0,1\}^n} \ket{z} .
\label{eq:s-def}\ee

\subsection{Quantum Approximate Optimization Algorithm (\QApx{})}
This algorithm is designed to find a string $z$ that approximately
maximizes $C$.  Let
\be C_{\max}:=\max_{z\in\{0,1\}^n} C(z)\label{eq:MC-obj}.\ee
Then we seek a $z$ such that the approximation ratio defined as
${C(z)}/{C_{\max}}$ is large.  The algorithm depends on an integer
$p \geq 1$ and we begin by describing the $p=1$ version.  Here we start with
two angles $\gamma$ and $\beta$ and use the quantum computer to construct
the state
\be \ket{\gamma, \beta}  = \exp(-i\beta B)\exp(-i\gamma C) \ket s.
\label{eq:QAOA-state}\ee
The goal now is to make \be \bra{\gamma,\beta} C \ket {\gamma,
  \beta}\label{eq:QAOA-obj}\ee as big as possible.  One way to do this
is to search for the $\gamma$ and $\beta$ which make \eq{QAOA-obj}
big.  For a fixed $\gamma$ and $\beta$, use the quantum computer to
make the state $\ket{\gamma,\beta}$ and then measure in the
computational basis to get a string $z$ and evaluate $C(z)$. Using an
enveloping classical search routine, hunt for the $\gamma$ and $\beta$
that maximize \eq{QAOA-obj} using the quantum computer as part of the
subroutine that produces values of $C$.  

However it was shown in
\cite{QAOA} that for a given  $\gamma$ and $\beta$
the value of \eq{QAOA-obj}
can be efficiently classically calculated.
  Once the best $\gamma$ and $\beta$ are chosen, the quantum computer would only be
used to find a string $z$ where $C(z)$ is equal to \eq{QAOA-obj} (or larger).  It was shown in
\cite{QAOA,QAOA2} that for certain combinatorial optimization problems
this algorithm can give approximation ratios that are better than what
can be achieved by randomly picking a string $z$.  The importance here
is that this is a quantum approximation algorithm with a provable
worst-case guarantee, something which had not been previously achieved
by any quantum algorithm for combinatorial optimization. However in
these cases the $p=1$ \QApx{} is outperformed by certain classical
algorithms that go beyond random guessing~\cite{random-CSP,GW95}.

Note that the circuit depth of the $p=1$ algorithm is low.  If the
cost function $C$ is a sum of $m$ terms each of which say involves 2
qubits as in \eq{disagree} then the first unitary hitting $\ket s$  in \eq{QAOA-state} can be
written as the product of $m$ commuting two-qubit unitaries.  The
instance-independent $B$
term is
a sum of $n$ commuting one-qubit operators.  The 
low circuit depth and simple form make the $p=1$ \QApx{} an attractive
candidate to run on a near-term gate-model computer. 

For $p > 1$ we select $2p$ angles $\vec\gamma := (\gamma_1, \ldots,\gamma_p)$ and
$\vec\beta := (\beta_1,\ldots \beta_p)$ and construct the state
\be |\vec\gamma , \vec\beta\rangle = 
 \exp(-i \beta_p B) \exp(-i\gamma_p C) \cdots 
\exp(-i \beta_1 B)\exp(-i\gamma_1 C) \ket{s}.
\label{eq:QAOA-state-p}\ee
The goal is now to find the $2p$ angles  $\vec\gamma$ and  $\vec\beta$ that make 
\be \bra{\vec\gamma , \vec\beta} C\ket{\vec\gamma , \vec\beta}
\label{eq:QAOA-obj-p}\ee
as big as possible.  For a given $\vec\gamma,\vec\beta$, we can use a quantum computer to evaluate \eq{QAOA-obj-p}.  This can be used as a subroutine in an enveloping classical algorithm to find the best angles $\vec\gamma,\vec\beta$.
 Note that performance can only improve as $p$
increases because the maximization over angles at level $p-1$ is a
constrained version of the maximization at level $p$ where the last two
angles are set to 0.  In fact it was shown in \cite{QAOA} that as $p$ goes to infinity, $C_{\text{max}}$ can be achieved.

As in the $p=1$ case, classical computers could be used to calculate \eq{QAOA-obj-p}, but now with a run-time that grows doubly exponentially in $p$~\cite{QAOA}.  Even for small $p$ this classical calculation can be impractical. 
By contrast, the quantum circuit depth is $p$ times the circuit
depth of the $p=1$ circuit.  
For small $p$ it might be possible to run this algorithm on a near-term gate-model quantum computer and explore
its performance empirically.


\subsection{The Polynomial Hierarchy}\label{sec:complexity}

Here we offer a review of a few concepts from complexity theory needed
to understand the results of this paper.  We will discuss the
polynomial hierarchy ($\PH$) and what it means to say the the $\PH$
collapses.  
We will also discuss the relationship between counting problems and
the $\PH$.  For a fuller account see~\cite{Arora-Barak}.


We begin by reviewing what the complexity class $\NP$ is by starting with
the familiar case of 3SAT.  3SAT is an example of a clause-based
decision problem over $n$-bit strings with each clause acting only on a
subset of the bits of size $\leq 3$.  The input is a string
$c$ that encodes the clauses and we ask if there exists an $n$-bit string $z$
that satisfies all of the clauses. The key feature of being in $\NP$
is that if there is a satisfying assignment $z$ and you have $z$ in
hand then you can quickly
check that all of the clauses are True on the string $z$.  The
satisfying string is called a ``proof'' or a ``witness'' because it can be
used to establish that the instance is indeed satisfiable.  On the
other hand, if an instance has no satisfying assignment then there is
no known witness
that can in general convince you of that in short order.   

We can think of $\NP$ as the set of problems specified by an input $c$
(encoding clauses in the case of 3SAT)
 which acts on strings $z$ through an easily computable
function $\eval(c,z)$, defined by
\be \eval(c,z) = 
\begin{cases}  
1 & z \text{ satisfies }c \\
0 & z \text{ does not satisfy }c\;.
\end{cases}\label{eq:check-def}\ee
The function $\eval(c,z)$ depends on the problem at hand whose clauses are encoded as $c$.
For a general combinatorial combinatorial search problem with cost
function \eq{CSP-def} we have that $\eval(c,z)=1$ iff
$C(z)=m$, where $C(z)$ is the number of clauses in $c$ satisfied by $z$.
Then the decision question becomes ``Given an input $c$, is the following statement true?'':
\be \text{There exists } z \text{ such that }\eval(c,z) = 1.
\ee

One way to solve the decision problem is to look through all $2^n$
strings $z$ to see if one or more satisfies the clauses.  It is
generally believed that there is no efficient (that is, polynomial in
$n$) algorithm which can answer the decision question for all problems
in $\NP$.  Equivalently it is believed that no efficient algorithm exists
for 3SAT.  This belief is written as $\Ptime\neq \NP$ where
$\Ptime$ is the class of decision problems that can be solved in
polynomial time. 

The polynomial hierarchy is a tower of classes of problems which
seemingly require more and more resources to solve.  At level 0 we
have the class $\Ptime$ and at level 1 we have the class $\NP$.  At level 2
we have problems specified by clauses encoded as $c$ which act on two
 $n$-bit strings $z_1$ and $z_2$.  Again there is an efficiently
computable function  $\eval(c,z_1,z_2)$ such that 
\be 
\eval(c,z_1,z_2)=
\begin{cases}  1 &  z_1,z_2 \text{ satisfies }c \\
0  & z_1,z_2 \text{ does not satisfy }c\;.
\end{cases}\label{eq:check-2}\ee
Now the decision question becomes ``Given an input $c$, is the following statement true?''
\be \text{For all }z_1, \text{there exists } z_2 \text{ such that } \eval(c,z_1,z_2)=1.
\label{eq:level-2}\ee

We can see that this is (apparently) exponentially harder than $\NP$ as follows.
Suppose we have a magical machine or ``oracle'' that can solve $\NP$ in
a single step.  
For a fixed $c,z_1$ we can view $\eval(c,z_1,z_2)$ as an efficiently computable
function with clauses $c,z_1$ acting on strings $z_2$. 
  For fixed
$c,z_1$ we give this function to the oracle which can quickly say if
there is a satisfying string $z_2$.   However to answer the decision
question at level 2 we may need to cycle through all $2^n$ values of
$z_1$.  So here the ability to solve quickly the level-1 problem does
not give rise to an efficient solver at level 2.   The argument here
is similar to the argument that $\Ptime$ is unlikely to equal $\NP$
since brute-force enumeration of witnesses takes exponential
time. It is only a conjecture that level 2 of the polynomial
hierarchy is exponentially harder than level 1, but this conjecture is
widely believed for the same reasons that $\Ptime\neq\NP$ is
widely believed.

At the next level of the hierarchy, level 3, we have clauses which act
on strings $z_1,z_2,z_3$ and the decision question becomes ``Given $c$,
is the following statement true?''
\be \text{There exists } z_1, \text{ such that for all }z_2, \text{ there exists } z_3\text{ such that }   \eval(c,z_1,z_2,z_3) = 1.\ee
Again we can see that an oracle at level 2 may need to be
called an exponential number of times to solve the decision question
at level 3, so in that sense level 3 is harder than level 2.  Hopefully
it is now straightforward to see how higher levels of the hierarchy
are defined and that each level is exponentially harder than the level
below.  The $\PH$ is the union of these decision problems at all levels. 

One way to understand the levels of the hierarchy is to consider the
game of chess where we can consider an analogous hierarchy where we
see the alternating quantifiers (but not the scaling with $n$).  Suppose
we ask ``For a board in position $c$, does White have a guaranteed
mate in $k$ moves?''.  For example at level 3 of the chess hierarchy
we are asking if there exists a move for White, $z_1$, such that all
of Black's moves $z_2$ can be countered with a move $z_3$ with which
White mates Black.

In this paper we are going to say that ``something'' implies the
collapse of the $\PH$ so we now want to say what the collapse of the $\PH$
means.  The ultimate collapse of the $\PH$ would come about if P=$\NP$
which would automatically imply that all levels of the hierarchy are
equal to P.  However we are only going to be able to show that
“something” implies the collapse of the hierarchy at level 3. What
this collapse means is that given an oracle for the level-3 decision
problem, that oracle can be called only a polynomial number of times
to solve the level-4 decision problem.  This would imply that all
levels of the hierarchy above 3 are equal to level 3 but it would not
imply that $\Ptime=\NP$.  Still it seems very implausible that the
hierarchy collapses at any level.  One could object here and say that
presuming the existence of an oracle at level 3 is unreasonable so we
shouldn't care what it would imply.  Aren't we saying that if dogs could talk
then pigs could fly?  No.  We are saying that no matter what the
computational resources are of the level-3 oracle, that is
unreasonable to suppose that a
polynomial number of calls to that oracle would solve the level-4
problem.

\subsection{Counting}

Return to our example of 3SAT and now ask the question: how many
strings $z$ satisfy an instance with clauses c?  This is not a
decision problem so it does not sit in the class $\NP$.  But if we had
an oracle which could solve this problem, in other words an efficient
function of $c$ which counts the number of satisfying assignments then
we could solve the decision problem by simply seeing if the answer is
0 or not.  In this sense the counting problem is harder than the
decision problem. The class of counting problems of this nature is
called $\SharpP$. This class includes problems such as finding the
permanent of a matrix whose entries are 0's and 1's. In fact the
$\SharpP$ class can be viewed (somewhat imprecisely) as containing the
whole $\PH$.  More precisely, any problem in $\PH$ can be solved with
a poly-time classical computer that can make a polynomial number of
calls to a $\SharpP$ oracle.

\section{Efficient Classical Computation of Matrix Elements of Quantum Circuits Implies that $\Ptime=\NP$}\label{sec:mat-elt}

Here we  give an argument for why exactly computing matrix elements of
quantum circuits is hard based on beliefs from complexity theory that
we just reviewed.  Our argument will then be slightly modified to show
that even computing the matrix elements of a $p=1$ \QApx{} circuit is hard.

We consider a quantum computer that can
implement gates of the form $\expp{-i\gamma C}$, where $C(z)$ arises from
a CSP with $m$ clauses as described in \eq{CSP-def}.   The quantum computer can also
implement the Hadamard gate on each qubit, meaning the gate
$$H  = \frac{1}{\sqrt{2}}\bpm 1 & 1 \\ 1 & -1 \epm.$$
  For each 
$r\in\{0,1,\ldots,m\}$, consider the matrix
element
\be \bra{0^n} H^{\ot n} \expb{-\frac{2\pi i r}{m+1}C}H^{\ot n}\ket{0^n},
\label{eq:matrix-elt}\ee
which is equivalently (using the definition of $\ket s$ from \eq{s-def})
\ba
=&\bra{s} \expb{-\frac{2\pi i r}{m+1}C} \ket{s} \label{eq:matrix-elt2}\\
=& \frac{1}{2^n} \sum_{z\in \{0,1\}^n}\expb{-\frac{2\pi i r}{m+1}C(z)}.\\
\intertext{
If we define $p_v$ to be the fraction of $z$ for which $C(z)=v$, then
\eq{matrix-elt} is}
 =& \sum_v p_v \expb{-\frac{2\pi i r}{m+1}v}.
\ea
This is equal to the Fourier transform of $p_v$.  Thus if we can
compute the matrix element \eq{matrix-elt} for each value of $r$ then
we can reconstruct the distribution $p_v$ by performing the inverse
Fourier transform on the $m+1$ amplitudes.  Note that we can assume that
$m+1$ is of order $n$ so that this operation can be done efficiently in
$n$.  Now knowing the exact value of $p_m$ would let us compute how many strings satisfy
all of the clauses, a $\SharpP$-hard problem.
As we discussed in the last section if this
could be calculated efficiently then we would have $\Ptime=\NP$.
Since we believe that $\Ptime\neq\NP$ we conclude that the matrix
elements \eq{matrix-elt} can not be efficiently calculated on a
classical computer.  We view this argument as a concrete version of
Feynman's intuition.


We now specialize to the case of the \QApx{}. The $p=1$ \QApx{} with
$\beta=0$ produces the state  
\be \expb{-\frac{2 \pi i r}{m+1} C} \ket{s} .\ee
and  if we take the $\ket{s}$  component we get \eq{matrix-elt2} which is \eq{matrix-elt}.
This means that an efficient classical algorithm for computing this
particular matrix element of the output of the \QApx{} would imply that
$\Ptime=\NP$. 

We have just argued that a classical computer cannot (under
reasonable assumptions) exactly compute matrix elements of general
quantum circuits, or even of the lowest depth \QApx{}.  But quantum
computers do not output matrix 
elements. Rather a quantum circuit produces a state which is then
measured in some basis.  The outcome of the measurement will follow
the probability distribution arising from the amplitudes squared but
you cannot decide in advance which outcome will be obtained.  So a
question that we can ask is ``Can a classical computer which uses
random bits, produce outcomes which follow the same probability
distribution that a quantum circuit would give?''  Later we will
show that under reasonable assumptions the distribution of outcomes of
even the $p=1$ \QApx{} cannot be faithfully reproduced by a classical
computer.

\section{Post-Selected Quantum Computing}
\label{sec:PostBQP}

In this section we will describe a model of quantum computing that
allows one to decide in advance what the outcome of a measurement will
be.  This model does not correspond to anything we imagine being able
to do with a real quantum computer but it will serve as a stepping
stone to establishing other results.   First we see that with this
model we can solve the Grover problem with one call. The search space
has size $N$ and there are $M$ marked items. The function $f(z)$ is  equal
to 1 if $z$ is a marked item and 0 otherwise.  Start with the uniform
superposition of $N$ basis vectors with an additional one-qubit register
set to 0:
\be \sum_{z=1}^N  \frac{1}{\sqrt{N}}  \ket z \ket 0,\ee
and then compute $f(z)$ into the second register so we have
\be \sum_{z=1}^N  \frac{1}{\sqrt{N}}  \ket z \ket{f(z)}.\ee
Now if we post-select on discovering upon measurement that the second register is a 1 we get
\be\sum_{z\text{ such that } f(z)=1}  \frac{1}{\sqrt{M}}  \ket{z}, \ee
where $M$ is the number of marked items.  So we see that with
post-selected quantum computing we could solve the Grover problem with
one call.   This means that with a polynomial-time post-selected quantum computer we could also solve an $\NP$-complete problem
such as 3SAT.

Post-selected quantum computing has even more computational power in
that it allows one to solve counting problems.  There is a result due
to Aaronson~\cite{Aaronson04} that says that the ability to do
post-selected quantum computing allows one to solve problems in $\SharpP$,
that is, to perform exact counting.  To see why this is so let us use
post-selection to count the number of marked items $M$ in the Grover
problem.  Start with uniform superposition $\ket s$  as we just did and
append another register which is in the state $\ket +$.
$$\sum_{z=1}^N \frac{1}{\sqrt{N}} \ket z  \ket + $$
where					
$$ \ket + = \frac{\ket 0 + \ket 1}{\sqrt 2}\,,$$
and act with
\be (-1)^{ \bigl [f \ot \proj{1}\bigr]}\;,\ee
where $f$ is the operator, acting on the first register, corresponding to the function $f(z)$.  The state now, up to normalization, is
\be \sum_z \ket z \ket 0
+ \sum_z (-1)^{f(z)} \ket z \ket 1.\ee
Now we measure the first register post-selecting on the outcome being
$\ket s$ .   This resulting state is, up to normalization,
\be \ket 0  + ( 1 - 2  M / N) \ket 1.\label{eq:before}\ee
Perform a Hadamard on this bit and get a state proportional to
\be (N-M)\ket 0 + M\ket 1.
\label{eq:blatz}\ee

For illustration suppose we know that $M$ is $N/2$ or $N/2+1$ and
we want to figure out which case we have.  Since we imagine that $N$ is
exponentially big, the two coefficients in \eq{blatz} differ from each other only by an
exponentially small amount so even multiple copies of this state will
not allow us to tell which one of the two coefficients is bigger.  But
we can use post-selection again.  

Even without post-selection we can use the old idea of {\em
  unambiguous state discrimination}~\cite{unambiguous}.  This is a
method for distinguishing any two distinct quantum states with zero
probability of error, but with some probability of outputting ``don't
know.''  Suppose the states are $\ket{\psi_1},\ket{\psi_2}\in \bbC^2$.
Let $\ket{\psi_2^\perp}$ be the unique vector (up to
phase) that is orthogonal to $\ket{\psi_2}$, and similarly for $\ket{\psi_1^\perp}$.
One way to perform unambiguous state discrimination is randomly either measure in the
$\{\ket{\psi_1},\ket{\psi_1^\perp}\}$ basis or the
$\{\ket{\psi_2},\ket{\psi_2^\perp}\}$ basis.  If the outcome is $\ket{\psi_1^\perp}$ then
we know that the original state must have been $\ket{\psi_2}$ and likewise if the outcome is
$\ket{\psi_2^\perp}$ then
we know that the original state must have been $\ket{\psi_1}$.  However, if the outcome is
$\ket{\psi_1}$ or $\ket{\psi_2}$ then we cannot definitively determine the original state and
so we output ``don't know''.  The probability of the ``don't know'' outcome is
$\frac 12 + \frac 12 |\braket{\psi_1}{\psi_2}|^2$.  This is always $<1$ unless $\ket{\psi_1}$ and
$\ket{\psi_2}$ are identical up to phase. (While the ``don't know'' probability can be
made lower~\cite{unambiguous}, this will not be necessary for our argument.)
 If the states are very close but not equal, say if they
correspond to $M=N/2$ and $M=N/2+1$, then the probability of ``don't
know'' will be very close, but not equal to, 1.  In this case, we can
post-select on getting an answer other than ``don't know'' and
the result will be correct with probability 1.

This idea can be generalized to distinguish between the cases
$M\leq N/2$ and $M>N/2$, as described by Aaronson in
\cite{Aaronson04}.  The new complication is that we don't know how far
$M$ is from $N/2$.  If, say, $M>\frac 34 N$ or $M < \frac 14 N$ then
we can distinguish these cases even without the help of
postselection.  Otherwise it is possible to use postselection to
amplify the difference between these two possibilities.  We can
unitarily map the state \eq{before} to 
\be \ket{0}\ot\frac{\ket 0 + \sqrt{3}\ket 1}{2} + (1-2M/N)\ket
{1}\ot\ket 0.\ee
If we then postselect on the second qubit being in the $\ket 0$ state
we will be left with a state proportional to
\be \ket 0 + 2(1-2M/N)\ket 1.\ee
Now performing a Hadamard yields a state proportional to
\be (N-M')\ket 0 + M'\ket 1,\ee
where $M'-N/2 = 2(M-N/2)$.  In other words we have doubled the
difference between $M$ and $N/2$.  If now $M'>\frac 34 N$ or $M' <
\frac 14 N$ then states can be easily distinguished, or if not then we
can continue amplifying.  This whole process requires only $O(\log N)$
time and copies of the original state, and so can be done efficiently
by a quantum computer with post-selection. 

Now we further extend the algorithm.  The same argument that worked
for $N/2$ could work for any threshold.  Repeatedly calling this
subroutine can then allow a postselected
quantum computers to determine the exact value of $M$, i.e.~the
number of satisfying assignments for a combinatorial search
problem such as 3SAT.

The complexity class $\BQP$, Bounded-error Quantum Polynomial time, is what
we generally think of as conventional quantum computing with a
polynomial number of qubits and a polynomial number of unitary
gates. Technically it is a class of decision problems which means that
we are using the quantum computer to answer Yes-No questions.  But
sometimes people say that the Shor factoring algorithm is in $\BQP$
because we could turn factoring into a decision problem (``is there a
factor less than $F$?'') and we will not bother too much with this kind
of distinction.  What we mean by $\PostBQP$ is what we have running a
quantum computer with a polynomial number of qubits and a polynomial
number of gates with the additional magical power of being able to
post-select on the outcome when a subset of the qubits are
measured. We have seen that $\PostBQP$ allows one to solve $\SharpP$ problems
and in fact (loosely speaking) it contains the whole $\PH$.

The classical analog of $\BQP$ is called $\BPP$ and it is the class of
decision problems that can be answered with bounded error in
polynomial time allowing for randomness as part of the algorithm.  So
$\BPP$ is what we think of as what can be achieved with conventional
classical computers.  Now $\PostBPP$ allows us to post-select on some
subset of the bits after the circuit is run.  For example, consider
the Grover problem where the conventional computer has two registers,
$z$ for the input and second register where $f(z)$ is stored.  Now if we
pick $z$ at random and post-select on the second register being 1, 
we find a marked item.
 This means that $\PostBPP$ has even more power than $\NP$.   Even though $\PostBPP$ is
very powerful, it is not believed to be powerful enough to solve
counting problems.  In fact $\PostBPP$ is known to be contained in the
third level of the $\PH$~\cite{postBPP}.   (We are not going to try to explain this
as it would take us too far afield.)
As a result, if $\PostBPP$ were to equal
$\PostBQP$ then the $\PH$ would collapse to the third level; meaning
that level 3 of the $\PH$ would equal level $k$ for all $k\geq 3$.
This statement is a key fact from complexity theory that we need for what follows.

In our proof that $\PostBQP$ can exactly compute $M$, there is only one step that could not be performed in  $\PostBPP$. This is the Hadamard mapping \eq{before} to \eq{blatz}.  In this way
quantum computers with postselection can amplify the distance of $M$
from $N/2$, while classical computers with postselection could only
amplify the distance of $M$ from 0 or $N$.  This is enough for
postselected classical computers to perform approximate counting but
(apparently) not exact counting.

\section{Efficient Classical Sampling of the Output of an Arbitrary Quantum Circuit
Implies the Collapse of the Polynomial Hierarchy}
\label{sec:arb-collapse}

Suppose you have a polynomial-size quantum circuit which produces the
state   $U\ket 0$ and you measure in the $z$ basis.  The probability of getting $z$ is
\be q(z) = \big|\bra{z} U \ket{0}\big|^2.\ee
The quantum computer will naturally produce strings $z$ with this
probability.  Now suppose that you had a classical device that uses
random bits and produces strings z with probability $p(z)$.   What we
argue in this section is that if $p(z)$ is close to $q(z)$ for all
poly-size quantum circuits $U$ and all inputs $z$ then
$\PostBPP = \PostBQP$.   But this would imply the collapse of the
$\PH$  as we just explained at the end of the last section.
For this reason it is believed that classical computers cannot
simulate the outputs of general quantum computers.   This is all 
known and our goal here is to familiarize the reader with these
arguments since they are necessary background for our
statements below about sampling from the output of the \QApx{}.

Since classical computation is a subset of quantum computation we have
that $\BPP\subseteq \BQP$.  Suppose that with a classical computer
using random bits we could efficiently produce strings with
probability $p(z)$ with $p(z)$ very close to $q(z)$.  
This means that the
classical device would have the same power as $\BQP$ so in this case we
would have that $\BPP=\BQP$.  

We now show that if we could classically efficiently sample from the
output of a quantum computer, then $\PostBPP=\PostBQP$.
First note that $\PostBPP\subseteq \PostBQP$  for the same reason that
$\BPP\subseteq \BQP$.  Now we need to show that if a 
decision problem can be solved with a post-selection on a quantum
computer and quantum computer outputs can be classically sampled then
the same decision problem can be solved with a classical computer that
uses post-selection; in other words, that $\PostBQP \subseteq \PostBPP$.

Consider a general quantum circuit $U$ acting on a tensor product
Hilbert space where the first factor is a single qubit and the second
is an $n$-qubit space. Let 
\be q(z_1, z_2)  =  \big| \bra{z_1 , z_2} U \ket{0 , 0^n} \big| ^2.\ee
We are going to post select on $z_2 = 0^n$ so we define
\be q_\text{post}(z_1)= \frac{q(z_1, 0^n)}{q( 0, 0^n) + q( 1, 0^n)},\ee
which is our normalized post-selected probability distribution over
the outcomes $z_1=0,1$.
A decision problem (one with a YES/NO answer) specified by some input
is said to be in $\PostBQP$ if there is a quantum circuit $U$ depending on
the input such that $q_\text{post}(1) \geq 2/3$ in the 
 YES case and $q_\text{post}(1) \leq 1/3$ in the NO case.

Now suppose that we have a classical computer that using coin
flips can output strings $z_1,z_2$ with a probability $p(z_1,z_2)$ that is
very close to $q(z_1 , z_2)$.  If we post-select on the second string
being $0^n$ then we have 
\be p_{\text{post}}(z_1) = \frac{p(z_1, 0^n)}{p( 0, 0^n) + p( 1,
  0^n)},\ee
where again $z_1 \in \{0,1\}$.

Let us now describe more precisely our hypothesis that $p$ is close to $q$.
 Suppose that
\be 
\left | p(z_1,z_2) - q(z_1,z_2) \right| \leq 0.1\, q(z_1,z_2).\ee
These bounds on $p$ and $q$ immediately imply that $p_{\text{post}}$
and $q_{\text{post}}$ can be related as
\be \frac{0.9}{1.1} q_{\text{post}}(z_1)
\leq p_{\text{post}}(z_1)
\leq \frac{1.1}{0.9} q_{\text{post}}(z_1)
.\ee
This means that
$p_{\text{post}}(1)\geq 0.54$
 in the YES case and 
$p_{\text{post}}(1)\leq 0.41$ in the NO case, so the post-selected
classical computer can solve the decision problem; i.e.~the problem is
in $\PostBPP$.
We then see
that under the assumption of an efficient classical simulation of the
output of a general quantum computer we have that $\PostBQP \subseteq
\PostBPP$.  This would then imply 
$\PostBPP = \PostBQP$.

As we said at the end of the last section, if $\PostBPP = \PostBQP$. then
the $\PH$ collapses at the third level.  Since we believe this does not
happen we conclude that no classical device can efficiently produce
samples that match those output by a general quantum computer.  We now
argue that no classical device can efficiently produce samples that
match those output by the $p=1$ \QApx{}.  To this end we first show that
$\PostQAOA = \PostBQP$.

\section{$\PostQAOA = \PostBQP$}
\label{sec:post-equal}

It was shown by Bremner, Josza and Shephard~\cite{BJS10} that if a classical
computer could efficiently sample from the output of an IQP circuit
then the $\PH$ would collapse.  In fact IQP or ``Instantaneous Quantum
Polytime'' is rather similar in structure to the $p=1$ \QApx{}.  For IQP
there are two unitaries used to construct a circuit. The first is the
Hadamard 
$$
H= \frac{1}{\sqrt 2}
\begin{pmatrix} 1 & 1 \\ 1 & -1 
\end{pmatrix}$$
which acts on each individual qubit.  The second is a unitary diagonal in the $z$ basis of the form
\be U_D = \exp (i D)\label{eq:UC-def}\ee
where $D$ is a quadratic function of the $\sigma_z$ operators:
\be D= \sum_{k, l}  J_{kl} \sigma_z^k \sigma_z^l + \sum_k M_k \sigma_z^k \,,\label{eq:quadratic}\ee
with $J_{kl},M_k$ arbitrary coefficients.   The circuit is of the form
\be
 H^{\ot n} U_D H^{\ot n} \ket{0^n }
\label{eq:IQP}\ee
and measurement is made in the computational basis.  So we have
\be q^{\text{IQP}}(z) = \bigl| \bra{z} H^{\ot n} U_D H^{\ot n}  \ket{ 0^n} \bigr| ^2.\ee
What was shown in \cite{BJS10} is that if there exists an efficient
classical algorithm that can produce strings $z$ with a distribution
close to $q^{\text{IQP}}(z)$ then the PH collapses at the third level.  

We will now
essentially copy the argument of \cite{BJS10} with the $p=1$ \QApx{} replacing the IQP
circuit. 
To show the similarities between IQP and \QApx{}, let $\beta= \pi/4$ and
define for a single qubit 
$$\tilde{H} = \exp(-i\frac{\pi}{4}\sigma_x) =\frac{ 1}{\sqrt 2} 
\bpm 1 & -i \\ -i  & 1 \epm,$$
so we see that a  special case of the $p=1$  \QApx{} circuit 
\eq{QAOA-state} is
\be \tilde{H}^{\ot n} e^{-i\gamma C} H^{\ot n} \ket{0^n}.
\label{eq:tilde-QAOA}\ee
which is to be compared with \eq{IQP}.
Furthermore if the cost function $C$ is a sum of two-bit clauses then
$e^{-i\gamma C}$ is of the form \eq{UC-def} with $D$ of the form \eq{quadratic}.

\QApx{} is a special case of $\BQP$ so we have that
$\PostQAOA\subseteq \PostBQP$.  We now show the reverse inclusion, that $\PostBQP$ is contained
in $\PostQAOA$ by showing that any $\PostBQP$ circuit can be rewritten as a
$\PostQAOA$ circuit.  The tricky part is that $\BQP$ circuits can have
polynomial depth and we are going to use post selection to collapse a
general $\PostBQP$ circuit to the very shallow $\PostQAOA$.

Any $\BQP$ circuit can be rewritten in terms of a few basic unitary
gates.  One such universal gate set consists of $H$, $e^{i\frac{\pi}{8}\sigma_z}$ and the controlled-phase gate $e^{-i\frac{\pi}{4}(I-\sigma_z^1)(I-\sigma_z^2)}$~\cite{Aharonov03}.   If we set $\gamma=\pi/4$, then these latter two gates can be written (up to an overall phase) in the form $e^{-i\gamma C_a}$ for 0/1-valued clauses $C_a$:
\begsub{phase-gates}
 e^{i\frac{\pi}{8}\sigma_z} & = e^{i\frac{\pi}{8}} e^{-i\frac{\pi}{4}\proj 1} \label{eq:pi8}\\
e^{-i\frac{\pi}{4}(I-\sigma_z^1)(I-\sigma_z^2)}& = e^{-i\frac{\pi}{4}4\proj{11}} \label{eq:c-phase}
\endsub
Eq.~\eq{pi8} comes from a single one-bit clause whereas \eq{c-phase} requires repeating the same two-bit clause four times.

The crucial difference between a general quantum circuit and \eq{tilde-QAOA} is that in
\eq{tilde-QAOA} all the
diagonal gates are performed consecutively without any $H$'s interspersed, 
while in a general circuit these noncommuting gates are interspersed
without restriction. 
But if we introduce post selection we can
replace each internal $H$ gate with a gadget that involves an auxiliary
qubit and a post-selected measurement involving a circuit of the form
\eq{tilde-QAOA}.  To see how this works consider a general circuit
that consists of a sequence of one- and two-qubit gates of the form \eq{phase-gates}
interspersed with $H$ gates acting on individual qubits.  Consider one
internal $H$ acting say on qubit $j$ when the state of the system is
$$\ket\alpha_{j,\text{rest}}  = \ket{0}_j \ket a_{\text{rest}}  +  \ket{1}_j \ket b_{\text{rest}},$$
where $\ket{a}_{\text{rest}},\ket{b}_{\text{rest}}$ represent all qubits of the system other than
$j$.
Now in our post-selected \QApx{} circuit we add an auxiliary qubit which
is in the state  $H \ket{0}_{\text{aux}} = \ket +_{\text{aux}}$  since all qubits are initialized
in this state.   Act on the auxiliary qubit and qubit $j$ with the diagonal unitary
\be
   \bpm 1  \\ & i \\ & &  1\\ & & &  -i \epm. \label{eq:UC-special}\ee
This gate can also be written as
\be 
e^{-i\frac{\pi}{4} 6\proj{01}}\,
e^{-i\frac{\pi}{4} 2\proj{11}}
\ee
so we can see explicitly that it is of the form $e^{-i\frac{\pi}{4} C}$ with $C$ containing repeated clauses.
Next we act with 
$\tilde{H}$ on  qubit $j$, measure qubit $j$, and post select on the outcome
being 0.  It is straightforward to see that the remaining state is
$(H\ot I)\ket{\alpha}_{\text{aux},\text{rest}}$ where
$$\ket\alpha_{\text{aux},\text{rest}}  = \ket{0}_{\text{aux}} \ket a_{\text{rest}}  +
\ket{1}_{\text{aux}} \ket b_{\text{rest}}.$$
In this sense we have teleported the operation from
the qubit $j$ to the auxiliary qubit but using post selection
we guarantee that the measurement step gives the desired outcome. 

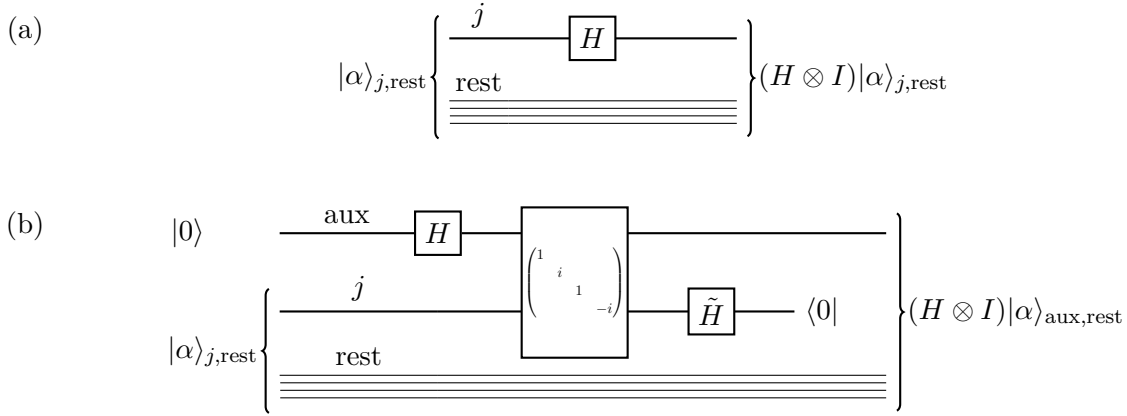
\begin{figure}[h]
\begin{tabular}{cc}
(a)\hspace{1cm} & 
    \begin{tikzpicture}[thick, baseline=(q2)]
    \tikzstyle{operator} = [draw,fill=white,minimum size=1.5em]

    \matrix[row sep=0.4cm, column sep=0.8cm] (circuit) {

&    \node (q2) {}; & \coordinate (q2a);&
    \node[operator] (H21) {$ H$}; &
    & \coordinate (end2);\\
  &  \node (q3) {}; &\coordinate (q3a);
     & &  &  \coordinate (end3); \\
    };
    \draw[decorate,decoration={brace},thick]
    ($(circuit.north east)-(0cm,0.2cm)$)
        to node[midway,right] (bracket) {$\displaystyle (H \ot I)\ket{\alpha}_{j,\text{rest}}$}
        ($(circuit.south east)+(0cm,-2mm)$);
        
        \draw[decorate,decoration={brace},thick]
        ($(q3)-(0mm,5mm)$)
        to node[midway,left] (bracket) {$\displaystyle \ket{\alpha}_{j,\text{rest}}$}
        ($(q2)+(0mm,3mm)$);

    \begin{pgfonlayer}{background}
        \draw[thick] (q2) to node[auto] {$j$} (q2a) (q2a) -- (end2);

        \draw[thin] (q3) to node[auto] {rest} (q3a);
        \draw[thin] ($(q3)+(1.5mm,-1mm)$) -- ($(q3a)-(0mm,1mm)$);
        \draw[thin] ($(q3)+(1.5mm,-2mm)$) -- ($(q3a)-(0mm,2mm)$);
        \draw[thin] ($(q3)+(1.5mm,-3mm)$) -- ($(q3a)-(0mm,3mm)$);
        \draw[thin] (q3a) -- (end3) ;
        \draw[thin] ($(q3a)+(0mm,-1mm)$) -- ($(end3)-(0mm,1mm)$);
        \draw[thin] ($(q3a)-(0mm,2mm)$) -- ($(end3)-(0mm,2mm)$);
        \draw[thin] ($(q3a)-(0mm,3mm)$) -- ($(end3)-(0mm,3mm)$);
    \end{pgfonlayer}
    \end{tikzpicture}

\\[2cm]
(b)\hspace{1cm} & 
    \begin{tikzpicture}[thick, baseline=(s1)]
    \tikzstyle{operator} = [draw,fill=white,minimum size=1.5em] 
    \tikzstyle{phase} = [draw,fill,shape=circle,minimum size=5pt,inner sep=0pt]
    \tikzstyle{operator2} = [draw,fill=white,inner sep=1pt,minimum height=2cm]

    \matrix[row sep=0.4cm, column sep=0.8cm] (circuit) {
    \node (s1) {\Ket{0}}; &[-2mm]
    \node (q1) {}; &[1cm]
    \node[operator] (H11) {$H$}; &[7mm]
&[7mm] &  &[-0.3cm]
    \coordinate (end1); \\
&    \node (q2) {}; & \coordinate (q2a);&
\coordinate (P22);
&    \node[operator] (H21) {$\tilde H$};
    & \node (end2) {$\bra 0$};\\
  &  \node (q3) {}; &\coordinate (q3a);
    & & &  &  \coordinate (end3); \\
    };
    \draw[decorate,decoration={brace},thick]
    ($(circuit.north east)-(0cm,0.2cm)$)
        to node[midway,right] (bracket) {$\displaystyle (H \ot I)\ket{\alpha}_{\text{aux},\text{rest}}$}
        ($(circuit.south east)+(0cm,-2mm)$);
        
        \draw[decorate,decoration={brace},thick]
        ($(q3)-(0mm,5mm)$)
        to node[midway,left] (bracket) {$\displaystyle \ket{\alpha}_{j,\text{rest}}$}
        ($(q2)+(0mm,3mm)$);

\node[operator2] at ($(P22)+(0,1em)$) {\scalebox{0.5}{$\bpm 1  \\ & i \\ & &  1\\ & & &  -i \epm$}}; 
    \begin{pgfonlayer}{background}
        \draw[thick] (q1) to node[auto] {aux} (H11);
        \draw[thick] (q2) to node[auto] {$j$} (q2a);
        \draw[thick] (H11) -- (end1) (q2a) -- (end2); 
        \draw[thin] (q3) to node[auto] {rest} (q3a);
        \draw[thin] ($(q3)+(1.5mm,-1mm)$) -- ($(q3a)-(0mm,1mm)$);
        \draw[thin] ($(q3)+(1.5mm,-2mm)$) -- ($(q3a)-(0mm,2mm)$);
        \draw[thin] ($(q3)+(1.5mm,-3mm)$) -- ($(q3a)-(0mm,3mm)$);
        \draw[thin] (q3a) -- (end3) ;
        \draw[thin] ($(q3a)+(0mm,-1mm)$) -- ($(end3)-(0mm,1mm)$);
        \draw[thin] ($(q3a)-(0mm,2mm)$) -- ($(end3)-(0mm,2mm)$);
        \draw[thin] ($(q3a)-(0mm,3mm)$) -- ($(end3)-(0mm,3mm)$);
    \end{pgfonlayer}
    \end{tikzpicture}
\end{tabular}
\caption{
(a) Somewhere inside of a big quantum circuit consisting of Hadamards and
diagonal unitaries, a Hadamard acts on qubit $j$.  The quantum state
before the Hadamard acts is $\ket\alpha$  which we denote as $\ket\alpha_{j, \text{rest}}$ to keep track of how it can be decomposed into qubit $j$ and
the rest.    This circuit element can be replaced by the post-selected
circuit in (b).  Qubit $j$ has been replaced by an
auxiliary qubit and the new circuit is of the form \eq{tilde-QAOA} with qubit $j$
post-selected to be $\ket{0}$.
}
\label{fig:post-select}
\end{figure}

We have just shown how to replace a Hadamard gate acting on a
qubit inside a general quantum circuit with a piece of a post-selected circuit of
the form \eq{tilde-QAOA}.  But we need to ensure that the whole
circuit is of the form \eq{tilde-QAOA}, meaning in addition that all the qubits in the
post-selected circuit start in the state $\ket 0$ and are then acted on
by a Hadamard and the last gate that any qubit is hit by is an $\tilde
H$.  We can ensure this as follows.
Suppose that in the original circuit, some qubit is first acted on by
a diagonal operator.  Then before the diagonal acts, insert $H\cdot H$ which
of course is the identity.  The first $H$ which acts maintains the form
\eq{tilde-QAOA}.  The second will need to be replaced by the construction in \fig{post-select}.
Now suppose that the last operator that hits some qubit is anything
other than $\tilde H$.  Then insert the gates $\tilde H \cdot \tilde H^\dag$ so the last
operator that acts on the qubit is indeed an $\tilde H$.  Now 
$$\tilde H^\dag =  H  \exp ( i \frac{\pi}{4}\sigma_z)  H$$
which explicitly is made from elements of our universal gate set.
Both $H$'s can be replaced by the construction of the previous
paragraph.  Thus, after all of these replacements the post-selected
circuit will be of the form \eq{tilde-QAOA}.

The construction just outlined allows us to rewrite any $\BQP$ circuit as
a post selected \QApx{} circuit.  So we have shown that $\BQP$ is
contained  in $\PostQAOA$.
But what about $\PostBQP$?  We wanted to show that this is contained
in $\PostQAOA$.
A $\PostBQP$ circuit will look like a $\BQP$ circuit except that at
the end some qubits will be post-selected onto the 0 state.  To
incorporate this into our $\PostQAOA$ simulation, we post-select
two batches of qubits into the 0 state: the ones that were
post-selected in the original $\PostBQP$ circuit and the new ones that
we used in our Hadamard gadgets.  The overall circuit still only uses
resources that are in $\PostQAOA$.

We have just shown that $\PostBQP \subseteq \PostQAOA$ and along with
the reverse inclusion we have that $\PostBQP =\PostQAOA$.
In fact our reduction does not introduce any error.  Suppose we have an
arbitrary quantum circuit built from Hadamards, $e^{i\frac{\pi}{8}Z}$
and controlled-phase gates whose post-selected output distribution is
$q_{\text{post}}(z)$.  
 We have shown that there exists a $C$ of
the form \eq{CSP-def} for which the corresponding $p=1$ \QApx{} circuit
satisfies
\be 
q^{\text{QAOA}}_{\text{post}}(z) = q_{\text{post}}(z).
\label{eq:post-QAOA-univ}\ee

\section{Efficient Classical Sampling of the Output of the $p=1$
  \QApx{} Implies the Collapse of the Polynomial Hierarchy}
\label{sec:QAOA-collapse}

Given an arbitrary \QApx{} circuit of the form \eq{tilde-QAOA} the probability distribution over
measurement outcomes $z$ is
\be 
q^{\text{QAOA}}(z) =
\left |\bra{z} \tilde H^{\ot n} e^{-i\frac{\pi}{4}C} H^{\ot n} \ket{0^n}\right|^2
\label{eq:qQAOA-def},\ee
where $C$ is of the form \eq{CSP-def}.
If we have a poly-time randomized classical algorithm that takes as
input $C$ of the form \eq{CSP-def} and outputs a string $z$
with probability $p(z)$ satisfying the bound
\be \left |p(z) - q^{\text{\QApx{}}}(z)\right| \leq
0.1 \, q^{\text{\QApx{}}}(z)
\label{eq:QAOA-acc}\ee
then $\PH$ collapses.   We now prove this by making use of the results
in the previous section.

Our strategy is to show that using the assumption of a classical simulator for \QApx{} satisfying \eq{QAOA-acc} we have $\PostBQP\subseteq \PostBPP$.  The only difference with \secref{arb-collapse} is that we have replaced general quantum circuits with \QApx{}.

Consider a problem in $\PostBQP$.   By definition there exists a quantum circuit $U$ with
output distribution $q(z_1,z_2)$ such that $q_{\text{post}}(1)\geq2/3$ if the
answer is YES and
$q_{\text{post}}(1)\leq 1/3$ if not.   This circuit can be assumed to be comprised of Hadamards and
two-qubit diagonal unitaries, since these are a universal gate set.
From the final result of the last section, there exists a
$p=1$ \QApx{} circuit with output distribution $q^{\text{QAOA}}(z)$
such that
$ q^{\text{\QApx{}}}_{\text{post}}(1) \geq 2/3$ in the YES case and
$ q^{\text{\QApx{}}}_{\text{post}}(1) \leq 1/3$ in the NO case.
Now using assumption \eq{QAOA-acc} and repeating the error analysis of
\secref{arb-collapse} we see that $p_{\text{post}}(1)\geq 0.54$ in the YES case
and $p_{\text{post}}(1)\leq 0.41$ in the NO case. 
  This means that a post-selected classical algorithm
could solve a problem in $\PostBQP$, proving that $\PostBQP\subseteq
\PostBPP$.  As we argued in \secref{arb-collapse} this implies the collapse of the
polynomial hierarchy since $\PostBPP$ is contained in the third level
of the $\PH$.  In other words, if we could efficiently classically sample from a
distribution $p(z)$ close to $ q^{\text{\QApx{}}}$ of \eq{qQAOA-def} with
tolerance \eq{QAOA-acc} then the $\PH$ collapses.

\mnb{Sampling vs estimating.} 
The QAOA circuit, for general $p$, produces states of the form \eq{QAOA-state-p}.   For
fixed $p$, if we specify $2p$ angles $\vec\gamma$ and $\vec\beta$ and
measure the state in the computational basis we find 
string $z$ with probability 
$$|\langle z | \vec\gamma,\vec\beta\rangle |^2.$$
It is this distribution that we claim (at least in worst case) cannot
be sampled from efficiently with a classical device. 

What if we only want to estimate
$\bra{\vec\gamma,\vec\beta}C\ket{\vec\gamma,\vec\beta}$?  It was shown
in \cite{QAOA} that this expected value can be determined on a
classical computer using resources that grow only polynomially in $n$
and $m$ but doubly exponentially in
$p$.
The situation is unusual in that we can classically determine
the expected value of the cost function, but we use the quantum
computer to produce strings with this value of the cost function.
The
distribution of the output is what cannot be reproduced classically.
(See also \cite{Aaronson14}.) 

From the optimization point of view, at least for low $p$, we can
classically determine the angles that maximize \eq{QAOA-obj-p} but we
use a quantum computer to find a string achieving roughly that value.
 Even if the QAOA has worse 
performance guarantees than certain classical algorithms, it can
exhibit Quantum Supremacy because its output distribution cannot be
efficiently classically sampled from. If for $p> 1$ or for
some problem at $p=1$, the QAOA outperforms all known classical
algorithms then it will achieve Quantum Supremacy in an algorithmic
sense.

\mnb{Sampling with additive error.}  Our results rule out efficient
classical sampling algorithm with low multiplicative error, as
specified in \eq{QAOA-acc}.  This requirement on the error is rather stringent.
Alternatively, one might consider ``additive error'' meaning that
$\sum_z |p(z)-q(z)|$ is upper bounded by a small constant.  It is more
challenging to rule out efficient classical simulation with additive error.
This can be done
for certain models of quantum computing~\cite{AA13,BMS15} but subject to
additional conjectures from complexity theory and/or probability
theory.  We believe that these results (especially \cite{BMS15}) could
be extended to cover the $p=1$ QAOA, but we leave this question
to future work.

\section{Quantum Adiabatic Algorithm}\label{sec:adiabatic}

We now discuss the prospects of sampling from the output of the
\QAdi{}.  We are thinking here of the \QAdi{} as being used to
solve optimization problems and not as a universal quantum computer.
We will place restrictions on the form of the Hamiltonian that governs
the evolution and see that these may make it more likely that
classical simulation is available.  To begin we review the basic idea
of the algorithm and set the notation.

\subsection{Definition and background}
 
With the Quantum Adiabatic Algorithm in optimization mode, we are
seeking the maximum of the cost function $C$ given by \eq{CSP-def}.
The basic building blocks are given by \eq{C-def}, \eq{B-def}, and
\eq{s-def}. 
This algorithm is designed to find the maximum of $C$, which is of
course the same as finding the minimum of $-C$, and we do this by
ground-state computation.  Note that $\ket{s}$ is the ground state of
$-B$ and we are seeking the ground state of $-C$.  Introduce a parameter-dependent Hamiltonian
\be H(s) = (1-s)(-B) + s (-C)
\qquad\text{with }0\leq s\leq 1
\label{eq:adia-Ht},\ee
a run time $T$, and a time-dependent Hamiltonian 
(which also depends on $T$):
\be \tilde{H}(t) = H( t/T).\ee
Then we evolve according to the Schr\"odinger equation
$$i \frac{d}{dt} \ket{\psi(t)} = \tilde H(t) \ket{\psi(t)}
\qquad\text{with}\qquad\ket{\psi(0)} = \ket{s}$$
which is the ground state of $\tilde H(0)$.  Run for time $T$ to get the
state $\ket{\psi(T)}$.  Note that the ground state of $\tilde H(T)$ is the
ground state of $-C$, so by the adiabatic theorem, in the $T\ra\infty$ limit the state $\ket{\psi(T)}$ is the ground state of $-C$.   

Actually for this to be true we need that the gap is not zero
throughout the evolution but this is guaranteed because 
with our choice of driving Hamiltonian, -B, (see \eq{B-def}),
the off-diagonal entries of $H(s)$ are non-positive and the Perron-Frobenius
theorem then guarantees a non-vanishing gap.  Hamiltonians with
real non-positive off-diagonal elements are called ``stoquastic'' and
this restriction is key to the ability to simulate ground-state properties of these
Hamiltonians.  

Note that the \QAdi{} will work perfectly if the run time is
infinite but we are interested in performance in realistic situations.
The required run time for good performance scales inversely with a
power of the minimum gap.  Here the community has struggled to analytically
derive useful bounds on the gaps of classes of instances.  
Much
numerical work has been done, enabled by the stoquastic nature of the
Hamiltonian; for example, see~\cite{FGHSSYZ}.  
The D-Wave device~\cite{DWave11,Dwave-tunneling16,DWave13,DWave-entanglement14} is designed to run the \QAdi{} and its performance
has been contrasted with simulations of the \QAdi{} as well as as with simulated annealing algorithms~\cite{DWave-100,DBIDBSMN,RWJBIWMLT}.
Here again the ultimate proof of performance may come
from running the algorithm on physical devices.

\subsection{Classical simulation of the Quantum Adiabatic Algorithm}
For the \QAdi{}, classical simulation may be easier than it is for
universal quantum computers. 
To argue this, we will need to restrict what we mean by adiabatic
quantum computation by assuming the following three conditions: 
\benum
\item $H(s)$ is stoquastic.  For simplicity we could assume that
  $H(s)$ is of the form in \eq{adia-Ht}. 
\item The gap between the bottom two eigenvalues of $H(s)$ is 
  $\geq 1/\poly(n)$ for all values of $s$.  
\item The total time $T$ is large enough that the adiabatic
  condition holds and the evolving state $\ket{\psi(t)}$ is
  approximately equal to the instantaneous ground state
  $\ket{g;s}$ (up to a phase) with $s=t/T$.   Because of our second
  assumption this means that $T= \poly(n)$ suffices.
\eenum
We call \QAdi{} subject to conditions 1-3 stoquastic gapped adiabatic
evolution, or \QAdiS{} for short.
When conditions 2 and 3 hold, they guarantee that the \QAdiS{} run on a quantum
computer will efficiently find the string minimizing the cost function $-C$.  Condition 1 is not necessary for the  success of the quantum algorithm but it will be needed for the classical simulation.
We discuss later the prospect of relaxing these  conditions.   

Since we are working with  $T$ large enough that the evolving state is
very close to the instantaneous ground state, we will approximate the
evolving state by the instantaneous ground state  $\ket{g;s}$  where
again $s = t/T$.  Now we are  interested in sampling in the
computational basis $\ket{z}$  from the ground state $\ket{g;s}$ and so the
string $z$ will be found with probability  $|\braket{z}{g;s} |^2$.  The
question now becomes: how hard is it to sample from this distribution?

The first thing to try is Quantum Monte Carlo, which is a family of
classical algorithms that can be used to sample from the ground states
of stoquastic Hamiltonians. (Please try not to be confused by the
terminology and remember that QMC is a classical algorithm used to
find properties of quantum systems.) If we choose an inverse
temperature $\beta$ that is sufficiently larger than the inverse gap of
$H(s)$ (meaning $\geq \poly(n)$, due to assumption 2 above) then
$$\frac{e^{-\beta H(s)}}{\tr e^{-\beta H(s)}}$$
will have high overlap with $\proj{g;s}$.
In this case 
  $|\braket{z}{g;s}|^2$ is approximately equal to
\be \frac{\bra{z}e^{-\beta H(s)}\ket{z}}{\tr e^{-\beta H(s)}}\,.
\label{eq:QMC-elt}\ee
Here we will use our stoquastic assumption which states that the
off-diagonal elements are non-positive.  In fact, let us also replace
$H(s)$ with $H(s)-cI$ for some $c\geq 0$ so that all entries of $H(s)$ are
non-positive and accordingly all entries of $-H(s)$ are non-negative.  This change will not affect the physics, nor will it
change \eq{QMC-elt}.
Now the numerator 
of \eq{QMC-elt} can be written as a
sum of an exponential number of nonnegative numbers.
There are various ways to do this but we will focus
specifically on an approach known as Path-Integral Monte Carlo.
For brevity, write $H=H(s)$ and note
\ba
\bra{z}e^{-\beta H}\ket{z} &=
\bra{z} (e^{-\frac{\beta H}{L}})^L\ket{z}
\\ &  =
\sum_{x_1 \in \{0,1\}^n}
\sum_{x_2 \in \{0,1\}^n}
\cdots
\sum_{x_L \in \{0,1\}^n}
\bra{z}e^{-\frac{\beta H}{L}}\ket{x_1}
\bra{x_1}e^{-\frac{\beta H}{L}}\ket{x_2}
\cdots
\bra{x_L}e^{-\frac{\beta H}{L}}\ket{z}.
\label{eq:PIMC}\ea
Each term $\bra{x_i}e^{-\frac{\beta H}{L}}\ket{x_{i+1}}$ is
nonnegative and, if $L$ is sufficiently large, can be 
approximated by expanding the exponential, although \eq{PIMC} as
written is exact.

Our goal now is to sample $z$ from a distribution proportional to \eq{PIMC}.  Let
\be w(z,x) :=  
\bra{z}e^{-\frac{\beta H}{L}}\ket{x_1}
\bra{x_1}e^{-\frac{\beta H}{L}}\ket{x_2}
\cdots
\bra{x_L}e^{-\frac{\beta H}{L}}\ket{z}
\label{eq:wzx-def}\ee
where $x$ denotes the tuple  $(x_1,\ldots, x_L)$.  Since $w(z,x)\geq
0$ we can define a normalized probability distribution 
\be p(z,x) := 
\frac{w(z,x)}{\sum_{z',x'}w(z',x')}
= \frac{w(z,x)}{\tr [e^{-\beta H}]}
.\ee
  If we can sample from strings $(z,x)$
according to $p$ then the marginal
distribution on $z$ (that is for fixed $z$ sum over $x$) will be equal to
\eq{QMC-elt} and with $\beta$  large enough this will be close to
  $|\braket{z}{g;s}|^2$.  So the task now is to sample from $p$.    To this end we use the
Metropolis method.  This method will produce a
Markov chain which settles down to following the distribution $p$.
Suppose that at some point in the process the configuration is $(z,x)$. 
The method requires
some rule for selecting possible next values of the configuration $(z',x')$.  For example it
might be to flip a bit (or some bits) at random.  The new value $(z',x')$ is accepted
according to a rule which only involves the ratio 
\be
\frac{p(z',x')}{p(z,x)}
=
\frac{w(z',x')}{w(z,x)}\,.
\ee 
For the problem at hand, using say single-bit
flips on $(z,x)$ we can easily compute the ratio of probabilities
knowing only $w$ which has
the explicit formula \eq{wzx-def}.  This means it is possible to sample from $p$ and then find out how
frequently the string $z$ would be measured in the ground state of
$H(s)$. 

What we just described will work in principle but it may take
exponentially long for the Markov chain to settle down to the desired
distribution.   To see that the process can take exponentially long to
settle consider a cost function $C$ for an $\NP$-complete problem with only
one string satisfying all the clauses. 
Then the ground state of $-C$
corresponds to this assignment and the gap of $-C$ is at least 1.
If $s=1$ and $\beta$ is large, then $e^{-\beta H(s)}$ has most of its
weight on the ground state, and so sampling from \eq{QMC-elt} will
have a high chance of revealing the satisfying assignment.  
It should not be possible for any algorithm, whether QMC or
anything else, to efficiently achieve this, unless $\NP$-complete problems could
be solved in polynomial time.
One might object that the $s=1$ case is too degenerate, since
$H(s=1)$ has no off-diagonal entries.  But the $s=1$ argument can be
extended to $s$ near 1, say $s= 1 - 1/\poly(n)$, to see that efficient
sampling from the ground state of this $H(s)$ would allow one to solve
$\NP$-complete problems.

One approach to attempt to sample from the ground state of $H(1)$ is
for the classical algorithm to track the adiabatic path.
We start with $s=0$, where the ground
state is the uniform superposition which we can easily sample from.
Then incrementally increase $s$ and for each value of $s$ run the
Metropolis procedure hopefully long enough to achieve the desired
distribution.  In this way we have ``warm starts'' for successive values
of $s$.  In fact this procedure can be viewed as a classical algorithm
for finding the minimum of $-C$ and is sometimes referred to as SQA for
Simulated Quantum Annealing~\cite{DWave-100,CrossonH16,FGHSSYZ,RWJBIWMLT}.


Despite the fact that there are cases (see the above discussion as well as Ref.~\cite{hastings-2013}) where we know that QMC takes exponentially long to equilibrate, QMC works well in practice in many cases.
In fact numerical simulations of \QAdiS{} out to 100's of qubits have been achieved.  More generally QMC is a standard tool in many-body quantum physics and has been successfully used to simulate quantum systems with millions of degrees of freedom by incorporating problem-specific insights.

\subsection{Formal Evidence that Stoquastic Gapped Adiabatic Evolution is Easier to Simulate than \QApx{}}

We can use QMC to provide formal evidence that \QAdiS{} is easier to
simulate than \QApx{}.  
Specifically, \QAdiS{}
can be simulated in $\PostBPP$; that is, by a poly-time classical randomized
computer with the ability to postselect.  Recall from
\secref{QAOA-collapse} that if \QApx{} could be simulated
in $\PostBPP$ then the
$\PH$ would collapse.   So, if the $\PH$ does not collapse then
\QAdiS{} can be thought of as easier to simulate then \QApx{}.
Another way to say this is that if we had an oracle for problems in
$\PostBPP$, we could call that oracle a polynomial number of times and
efficiently sample from the ground state of a gapped stoquastic
Hamiltonian, but we have strong reasons to believe  that such an
oracle would not allow for the efficient sampling of the \QApx{}. 

We now sketch the argument for why \QAdiS{} is in $\PostBPP$,
following \cite{bravyi-2006}.  This argument uses QMC but does not
rely on a Markov chain or any other method of sampling.  Instead
post-selection can be used to immediately jump to the right
distribution using rejection sampling.  
We know already that $w(z,x)\geq
0$, but from the form \eq{wzx-def} we can also bound $w(z,x)\leq
w_{\max}$ for some easily computable bound $w_{\max}$ which need not be tight.
These ingredients are now enough for a post-selected sampling
algorithm.   Let $b$ be a work bit that we will post-select on.  Then the algorithm is:
\benum
\item Choose $(z,x)$ uniformly at random. 
\item Set $b=1$ with probability $\frac{w(z,x)}{w_{\max}}$ and $b=0$
  with probability $1-\frac{w(z,x)}{w_{\max}}$.
\eenum
Conditioned on $b=1$ the resulting probability distribution over $(z,x)$
is exactly proportional to $w(z,x)$.  So if we post-select on $b=1$ we
get $(z,x)$ with probability $p(z,x)$.

As with many algorithms in $\PostBPP$, the probability of rejection
($b=0$) can be very high and so this does not give a polynomial time
algorithm for actual classical computers.  But this argument is still
enough to show that simulating \QAdiS{} can be done in $\PostBPP$.

\subsection{Universal adiabatic quantum computing}
It is known that adiabatic quantum computing is a universal form of
quantum computing~\cite{ADKLLR}.  This means that given any quantum
circuit with $n$ qubits and $\cT$ unitaries producing a state
$\ket\psi$, one can construct a local Hamiltonian $H(s)$ with gap
$\geq 1/\poly(n,\cT)$ that can be used in a poly-time adiabatic
algorithm to produce a good approximation of the same state
$\ket\psi$.  This construction makes use of a non-stoquastic
Hamiltonian to drive the evolution.

Suppose that \QAdiS{} could be used to perform
universal quantum computation.   We just showed that 
\QAdiS{} could be simulated in $\PostBPP$.  But this would imply
that $\BQP$ is in $\PostBPP$.  As we showed in \secref{arb-collapse} this would imply
that $\PostBQP$ is in $\PostBPP$, which in turn would collapse the
$\PH$.  So here again we see a crucial difference.  An oracle for
problems in $\PostBPP$ would allow the simulation of 
\QAdiS{} but the same oracle would not be strong enough to simulate
general \QAdi{}.

\section{Discussion}

We are entering an era when small-scale gate-model quantum computers
are being built.  The natural question is what algorithms should be
run on them?  Answering this question as the devices are being built
will influence the design of hardware architecture and help set
performance goals.

Devices exist and are being built to run the Quantum Adiabatic
Algorithm in optimization mode.  We argue that if the Hamiltonian
governing the evolution is stoquastic then simulation is more feasible
than it is for the \QApx{}.  We believe that this makes a compelling
case that future devices should be governed by Hamiltonians which are
not restricted to be stoquastic.  This will make simulation more
difficult both in practice and for complexity theoretic reasons. It
might be the case that a stoquastic system can be of real
computational value when compared with all known classical algorithms.
But still it would seem  prudent to operate quantum computers in a
way that  makes classical simulation more difficult.

Another candidate quantum algorithm to run on near-term quantum computers is the \QApx{}.  This gate model algorithm is
designed to find approximate solutions to combinatorial search
problems.   Running it may be competitive with or outperform
classical computers as the size of
quantum computers gets larger.  What we show in this paper is that
efficient sampling from the output of even the lowest depth version of
this algorithm would collapse the PH.  This means that, based on plausible conjectures from complexity theory, there are choices of $\gamma$, $\beta$ and cost function $C$ for which the output of a quantum computer running the $p=1$ \QApx{} could
not be mimicked with a classical device. This strengthens the case that
the \QApx{} should be run on a near term quantum device. 

\section*{Acknowledgments}

We are grateful to Michael Bremner, Daniel Brod, David Gosset, Sam
Gutmann, Jeffrey Goldstone and Eleanor Rieffel for helpful comments
and discussions.  Thanks to Alex Dalzell for catching a mistake in the
first version of this paper.
EF would like to thank the Google Quantum AI team for stimulating discussion.
EF was funded by NSF grant CCF-1218176. AWH was funded by NSF grants
CCF-1111382 and CCF-1452616 and IARPA 
via DoI/IBC contract
number D15PC00242.  Both were funded by ARO contract W911NF-12-1-0486.
The views and conclusions contained herein are those of the authors
and should not be interpreted as necessarily representing the official
policies or endorsements, either expressed or implied, of IARPA,
DoI/NBC, or the U.S. Government.


\begin{thebibliography}{10}

\bibitem{Aaronson04}
S.~Aaronson.
\newblock Quantum computing, postselection, and probabilistic polynomial-time.
\newblock {\em Proc. Roy. Soc. A}, 461:2063, 2005,
  \href{http://arxiv.org/abs/quant-ph/0412187}{{\ttfamily
  arXiv:quant-ph/0412187}}.

\bibitem{Aaronson14}
S.~Aaronson.
\newblock The equivalence of sampling and searching.
\newblock {\em Theor. Comp. Sys.}, 55(2):281--298, Aug. 2014,
  \href{http://arxiv.org/abs/1009.5104}{{\ttfamily arXiv:1009.5104}}.

\bibitem{AA13}
S.~Aaronson and A.~Arkhipov.
\newblock The computational complexity of linear optics.
\newblock {\em Theory of Computing}, 9(4):143--252, 2013,
  \href{http://arxiv.org/abs/1011.3245}{{\ttfamily arXiv:1011.3245}}.

\bibitem{Aharonov03}
D.~Aharonov.
\newblock A simple proof that {T}offoli and {H}adamard are quantum universal,
  2003,  \href{http://arxiv.org/abs/quant-ph/0301040}{{\ttfamily
  arXiv:quant-ph/0301040}}.

\bibitem{ADKLLR}
D.~Aharonov, W.~van Dam, J.~Kempe, Z.~Landau, S.~Lloyd, and O.~Regev.
\newblock Adiabatic quantum computation is equivalent to standard quantum
  computation.
\newblock {\em SIAM J. Comput.}, 37(1):166--194, Apr. 2007,
  \href{http://arxiv.org/abs/quant-ph/0405098}{{\ttfamily
  arXiv:quant-ph/0405098}}.

\bibitem{Arora-Barak}
S.~Arora and B.~Barak.
\newblock {\em Computational Complexity: A Modern Approach}.
\newblock Cambridge University Press, 2009.

\bibitem{random-CSP}
B.~Barak, A.~Moitra, R.~O'Donnell, P.~Raghavendra, O.~Regev, D.~Steurer,
  L.~Trevisan, A.~Vijayaraghavan, D.~Witmer, and J.~Wright.
\newblock {Beating the Random Assignment on Constraint Satisfaction Problems of
  Bounded Degree}.
\newblock In N.~Garg, K.~Jansen, A.~Rao, and J.~D.~P. Rolim, editors, {\em
  Approximation, Randomization, and Combinatorial Optimization. Algorithms and
  Techniques (APPROX/RANDOM 2015)}, volume~40 of {\em Leibniz International
  Proceedings in Informatics (LIPIcs)}, pages 110--123, Dagstuhl, Germany,
  2015. Schloss Dagstuhl--Leibniz-Zentrum fuer Informatik,
  \href{http://arxiv.org/abs/1505.03424}{{\ttfamily arXiv:1505.03424}}.

\bibitem{DWave-100}
S.~Boixo, T.~F. R{\o}nnow, S.~V. Isakov, Z.~Wang, D.~Wecker, D.~A. Lidar, J.~M.
  Martinis, and M.~Troyer.
\newblock Quantum annealing with more than one hundred qubits.
\newblock {\em Nature Phys.}, page 218, 2014,
  \href{http://arxiv.org/abs/1304.4595}{{\ttfamily arXiv:1304.4595}}.

\bibitem{Dwave-tunneling16}
S.~Boixo, V.~N. Smelyanskiy, A.~Shabani, S.~V. Isakov, M.~Dykman, V.~S.
  Denchev, M.~H. Amin, A.~Y. Smirnov, M.~Mohseni, and H.~Neven.
\newblock Computational multiqubit tunnelling in programmable quantum
  annealers.
\newblock {\em Nature Communications}, 7, 2016,
  \href{http://arxiv.org/abs/1502.05754}{{\ttfamily arXiv:1502.05754}}.

\bibitem{bravyi-2006}
S.~Bravyi, D.~P. DiVincenzo, R.~I. Oliveira, and B.~M. Terhal.
\newblock The complexity of stoquastic local hamiltonian problems.
\newblock {\em Quant. Inf. Comp.}, 8(5):0361--0385, 2006,
  \href{http://arxiv.org/abs/quant-ph/0606140}{{\ttfamily
  arXiv:quant-ph/0606140}}.

\bibitem{BJS10}
M.~J. Bremner, R.~Jozsa, and D.~J. Shepherd.
\newblock Classical simulation of commuting quantum computations implies
  collapse of the polynomial hierarchy.
\newblock {\em Proceedings of the Royal Society of London A: Mathematical,
  Physical and Engineering Sciences}, 467(2126):459--472, 2010,
  \href{http://arxiv.org/abs/1005.1407}{{\ttfamily arXiv:1005.1407}}.

\bibitem{BMS15}
M.~J. Bremner, A.~Montanaro, and D.~J. Shepherd.
\newblock Average-case complexity versus approximate simulation of commuting
  quantum computations.
\newblock {\em Phys. Rev. Lett.}, 117:080501, Aug 2016,
  \href{http://arxiv.org/abs/1504.07999}{{\ttfamily arXiv:1504.07999}}.

\bibitem{Brod15}
D.~J. Brod.
\newblock Complexity of simulating constant-depth {BosonSampling}.
\newblock {\em Phys. Rev. A}, 91:042316, Apr 2015,
  \href{http://arxiv.org/abs/1412.6788}{{\ttfamily arXiv:1412.6788}}.

\bibitem{unambiguous}
A.~Chefles.
\newblock Unambiguous discrimination between linearly independent quantum
  states.
\newblock {\em Physics Letters A}, 239(6):339 -- 347, 1998,
  \href{http://arxiv.org/abs/quant-ph/9807022}{{\ttfamily
  arXiv:quant-ph/9807022}}.

\bibitem{CrossonH16}
E.~Crosson and A.~W. Harrow.
\newblock Simulated quantum annealing can be exponentially faster than
  classical simulated annealing.
\newblock In {\em 2016 IEEE 57th Annual Symposium on Foundations of Computer
  Science (FOCS)}, pages 714--723, Oct 2016,
  \href{http://arxiv.org/abs/1601.03030}{{\ttfamily arXiv:1601.03030}}.

\bibitem{DBIDBSMN}
V.~S. Denchev, S.~Boixo, S.~V. Isakov, N.~Ding, R.~Babbush, V.~Smelyanskiy,
  J.~Martinis, and H.~Neven.
\newblock What is the computational value of finite range tunneling?, 2015,
  \href{http://arxiv.org/abs/1512.02206}{{\ttfamily arXiv:1512.02206}}.

\bibitem{DWave13}
N.~G. Dickson, M.~Johnson, M.~Amin, R.~Harris, F.~Altomare, A.~Berkley,
  P.~Bunyk, J.~Cai, E.~Chapple, P.~Chavez, et~al.
\newblock Thermally assisted quantum annealing of a 16-qubit problem.
\newblock {\em Nature communications}, 4:1903, 2013.

\bibitem{QAOA}
E.~Farhi, J.~Goldstone, and S.~Gutmann.
\newblock A quantum approximate optimization algorithm, 2014,
  \href{http://arxiv.org/abs/1411.4028}{{\ttfamily arXiv:1411.4028}}.

\bibitem{QAOA2}
E.~Farhi, J.~Goldstone, and S.~Gutmann.
\newblock A quantum approximate optimization algorithm applied to a bounded
  occurrence constraint problem.
\newblock Technical Report MIT-CTP/4628, MIT, 2014,
  \href{http://arxiv.org/abs/1412.6062}{{\ttfamily arXiv:1412.6062}}.

\bibitem{farhi00}
E.~Farhi, J.~Goldstone, S.~Gutmann, and M.~Sipser.
\newblock Quantum computation by adiabatic evolution.
\newblock Technical Report MIT-CTP-2936, MIT, 2000,
  \href{http://arxiv.org/abs/quant-ph/0001106}{{\ttfamily
  arXiv:quant-ph/0001106}}.

\bibitem{FGHSSYZ}
E.~Farhi, D.~Gosset, I.~Hen, A.~W. Sandvik, P.~Shor, A.~P. Young, and
  F.~Zamponi.
\newblock Performance of the quantum adiabatic algorithm on random instances of
  two optimization problems on regular hypergraphs.
\newblock {\em Phys. Rev. A}, 86:052334, Nov 2012,
  \href{http://arxiv.org/abs/1208.3757}{{\ttfamily arXiv:1208.3757}}.

\bibitem{FGHP}
S.~Fenner, F.~Green, S.~Homer, and R.~Pruim.
\newblock Determining acceptance possibility for a quantum computation is hard
  for the polynomial hierarchy.
\newblock In {\em Proceedings of the 6th Italian Conference on Theoretical
  Computer Science}, pages 241--252, 1998,
  \href{http://arxiv.org/abs/quant-ph/9812056}{{\ttfamily
  arXiv:quant-ph/9812056}}.

\bibitem{GW95}
M.~X. Goemans and D.~P. Williamson.
\newblock Improved approximation algorithms for maximum cut and satisfiability
  problems using semidefinite programming.
\newblock {\em JACM}, 42, 1995.

\bibitem{postBPP}
Y.~Han, L.~A. Hemaspaandra, and T.~Thierauf.
\newblock Threshold computation and cryptographic security.
\newblock In K.~W. Ng, P.~Raghavan, N.~V. Balasubramanian, and F.~Y.~L. Chin,
  editors, {\em Algorithms and Computation: 4th International Symposium, ISAAC
  '93 Hong Kong, December 15--17, 1993 Proceedings}, pages 230--239, Berlin,
  Heidelberg, 1993. Springer Berlin Heidelberg.

\bibitem{hastings-2013}
M.~B. Hastings.
\newblock Obstructions to classically simulating the quantum adiabatic
  algorithm.
\newblock {\em Quantum Information \& Computation}, 13(11-12):1038--1076, 2013,
   \href{http://arxiv.org/abs/1302.5733}{{\ttfamily arXiv:1302.5733}}.

\bibitem{DWave11}
M.~Johnson, M.~Amin, S.~Gildert, T.~Lanting, F.~Hamze, N.~Dickson, R.~Harris,
  A.~Berkley, J.~Johansson, P.~Bunyk, et~al.
\newblock Quantum annealing with manufactured spins.
\newblock {\em Nature}, 473(7346):194--198, 2011.

\bibitem{Nishimori:98a}
T.~Kadowaki and H.~Nishimori.
\newblock Quantum annealing in the transverse {I}sing model.
\newblock {\em Phys. Rev. E}, 58:5355--5363, Nov 1998,
  \href{http://arxiv.org/abs/cond-mat/9804280}{{\ttfamily
  arXiv:cond-mat/9804280}}.

\bibitem{DWave-entanglement14}
T.~Lanting, A.~J. Przybysz, A.~Y. Smirnov, F.~M. Spedalieri, M.~H. Amin, A.~J.
  Berkley, R.~Harris, F.~Altomare, S.~Boixo, P.~Bunyk, N.~Dickson, C.~Enderud,
  J.~P. Hilton, E.~Hoskinson, M.~W. Johnson, E.~Ladizinsky, N.~Ladizinsky,
  R.~Neufeld, T.~Oh, I.~Perminov, C.~Rich, M.~C. Thom, E.~Tolkacheva,
  S.~Uchaikin, A.~B. Wilson, and G.~Rose.
\newblock Entanglement in a quantum annealing processor.
\newblock {\em Phys. Rev. X}, 4:021041, May 2014,
  \href{http://arxiv.org/abs/1401.3500}{{\ttfamily arXiv:1401.3500}}.

\bibitem{Saeed16}
S.~Mehraban.
\newblock {C}omputational {C}omplexity of {S}ome {Q}uantum {T}heories in $1+1$
  {D}imensions, 2016,  \href{http://arxiv.org/abs/1512.09243}{{\ttfamily
  arXiv:1512.09243}}.

\bibitem{supremacy}
J.~Preskill.
\newblock Quantum computing and the entanglement frontier, 2012,
  \href{http://arxiv.org/abs/1203.5813}{{\ttfamily arXiv:1203.5813}}.

\bibitem{RWJBIWMLT}
T.~F. R{\o}nnow, Z.~Wang, J.~Job, S.~Boixo, S.~V. Isakov, D.~Wecker, J.~M.
  Martinis, D.~A. Lidar, and M.~Troyer.
\newblock Defining and detecting quantum speedup.
\newblock {\em Science}, 345(6195):420--424, 2014,
  \href{http://arxiv.org/abs/1401.2910}{{\ttfamily arXiv:1401.2910}}.

\bibitem{TD04}
B.~M. Terhal and D.~P. DiVincenzo.
\newblock Adaptive quantum computation, constant-depth quantum circuits and
  {A}rthur-{M}erlin games.
\newblock {\em Quantum Info. Comput.}, 4(2):134--145, Mar. 2004,
  \href{http://arxiv.org/abs/quant-ph/0205133}{{\ttfamily
  arXiv:quant-ph/0205133}}.

\end{thebibliography}
\end{document}